\documentclass[12pt,letterpaper]{article}

\usepackage{jheppub}
\usepackage{amsmath, amssymb}
\usepackage{verbatim}

\usepackage{graphicx}
\usepackage{subfigure}
\input{epsf}
\usepackage{epsfig}
\usepackage{epstopdf}
\usepackage{amsthm}
\usepackage{xcolor}
\usepackage{tcolorbox}

\usepackage{tikz}
\usepackage{tikz-feynman}

\def\({\left(} \def\){\right)}
\def\[{\left[} \def\]{\right]}
 
\def\del{{\partial}}

\newcommand{\non}{\nonumber \\}

\newcommand{\be}{\begin{equation}}
\newcommand{\ee}{\end{equation}}
\newcommand{\bea}{\begin{eqnarray}}
\newcommand{\eea}{\end{eqnarray}}
\newcommand{\ba}{\begin{eqnarray}}
\newcommand{\ea}{\end{eqnarray}}

\newcommand{\beq}{\begin{equation}}
\newcommand{\eeq}{\end{equation}}
\newcommand{\beqa}{\begin{eqnarray}}
\newcommand{\eeqa}{\end{eqnarray}}
\newcommand{\beqar}{\begin{eqnarray*}}
\newcommand{\eeqar}{\end{eqnarray*}}
\newcommand{\reef}[1]{(\ref{#1})}

\newcommand{\mt}[1]{\textrm{\tiny #1}}

\def\schw{ Schwarzschild }

\title{Gravito-magnetic Polarization of Schwarzschild Black Hole}

\author{Tomer Hadad,}
\author{Barak Kol}
\author{and Michael Smolkin}

\affiliation{The Racah Institute of Physics, The Hebrew University of Jerusalem, \\ Jerusalem 91904, Israel \\}

\emailAdd{tomer.hadad1@mail.huji.ac.il}
\emailAdd{barak.kol@mail.huji.ac.il}
\emailAdd{michael.smolkin@mail.huji.ac.il}

\abstract{We determine the gravito-magnetic Love numbers of non-rotating black holes in all spacetime dimensions through a novel and direct derivation. The Ishibashi-Kodama master field and its associated field equation are avoided. The matching to the EFT variables is simple. This method allows us to correct the values in the literature. Moreover, we highlight a parity-based selection rule for nonlinear terms that include both electric-type and magnetic-type gravitational field tensors, enabling us to conclude that many of the nonlinear response coefficients in the Schwarzschild black hole effective action vanish.}

\begin{document}
\maketitle

\section{Introduction}

Love numbers characterize the response of a celestial object to an external gravitational field \cite{Love1909}, namely a tidal deformation. More generally, they can refer to the response of any object to any external field. Love numbers of black holes are especially interesting, both for intrinsic theoretical reasons being fundamental properties of black holes and relativistic gravity (aka General Relativity or GR), and for observational reasons through their influence on the motion of inspiraling binaries, see e.g. \cite{Flanagan:2007ix,Cardoso:2017cfl,Cardoso:2019vof}. 

The first black hole Love numbers that were determined were those of the \schw black hole \cite{Damour:2009vw,Binnington:2009bb,Damour:2009b,Kol:2011vg}. It appears that this determination requires only the knowledge of the \schw metric \cite{Schwarzschild:1916uq}, which was available for almost a century earlier. However, the nonlinear nature of GR introduces an issue, namely nonlinear contributions appear to mix with the black hole response and thereby mask it.\footnote{This was stressed by \cite{Damour:2009vw,Damour:2009b}, see footnote 3 of \cite{Kol:2011vg}.}
This was overcome by dimensional regularization \cite{Kol:2011vg}. On the way, the Love numbers of static black holes in all dimensions, known as the Tangherlini solutions \cite{Tangherlini:1963bw}, were determined. Interestingly, the 4d BH Love numbers were found to vanish for all $l$, the spherical harmonic index. Another possible reason for the time that it took to determine the \schw Love numbers, is that perhaps the question was not asked.

Following the first determination of BH Love numbers numerous generalizations suggested themselves. Magnetic Love numbers correspond to the response to a gravito-magnetic external field. In 4d, they were evaluated already in \cite{Damour:2009vw,Binnington:2009bb}, while in arbitrary spacetime dimension, their value was stated more recently in \cite{Hui:2020xxx}. Additional generalizations include: non-gravitational Love numbers (the first such study might be that of scalar Love numbers in \cite{Kol:2011vg}), nonzero frequency Love numbers, Love numbers of more general black holes such as Kerr (see more below) or in higher dimensions, and nonlinear response coefficients \cite{DeLuca:2023mio,Riva:2023rcm}. Magnetic Love numbers are central to this work and the nonlinear ones will also be studied here. 

A second wave of progress on BH Love numbers started with the evaluation of Kerr Love numbers \cite{LeTiec:2020spy,Chia:2020yla}, following earlier work on slowly rotating black holes \cite{Poisson:2014gka,Pani:2015hfa,Pani:2015nua}. The wave progressed to proposals for a symmetry that is responsible for the vanishing of 4d Love numbers \cite{Charalambous:2021kcz,Hui:2021vcv,Hui:2022vbh,Charalambous:2022rre}, sometimes called Love symmetry.

An Effective Field Theory (EFT) perspective is implicit in the very notion of Love numbers, since by definition they allow to replace an object by a point particle equipped with some response coefficients. In field theory, such a procedure where the short distance scale of the object size is eliminated and replaced by effective couplings with the long distance scale of the external field, is known as an Effective Field Theory. In the case at hand, the EFT is called the effective point particle description of a black hole. In fact, this EFT is an ingredient of another EFT, the Non-Relativistic GR (NRGR) \cite{Goldberger:2004jt,Goldberger:2007hy,Kol:2007rx,Kol:2007bc} which presents the post-Newtonian approximation in the EFT framework.

In this paper, we examine two questions. The first question is to derive the magnetic Love numbers through the EFT perspective in all spacetime dimensions. \cite{Hui:2020xxx} studied these numbers by using the Ishibashi-Kodama equation \cite{Kodama:2003jz} and stated values for them.  As we shall see, it would turn out that not only do we get a novel and somewhat economized derivation for these numbers, but in fact, we correct a prefactor in the literature. The corrected value is now accepted by the authors of \cite{Hui:2020xxx} -- see its revised version. The nature of the economization is that both the Ishibashi-Kodama master field and the associated field equation are avoided, being replaced by a more direct, though not gauge-invariant, derivation, and the matching to the EFT variables becomes simpler.

The second question concerns nonlinear response coefficients. We shall point out a parity-based selection rule for nonlinear terms that include both the electric-type and magnetic-type gravitational field tensors. This means that many of the nonlinear response coefficients vanish, thereby allowing future work to concentrate on a considerably smaller set of such coefficients.

This work is organized as follows. In the Sect. \ref{sec:NRG}, we review the non-relativistic decomposition of the Einstein metric field, which is instrumental to this work. Sect. \ref{sec:MagLove} determines the magnetic Love numbers, starting with a GR analysis followed by an EFT matching. Sect. \ref{sec:nonlinear} describes and derives the nonlinear selection rule. Finally, a discussion and open questions are presented in Sect. \ref{sec:discussion}. Appendix \ref{Derivation of worldline operators} provides some details regarding the black hole point-particle effective action, while Appendix \ref{Evaluation gravito-A-7D-Diag} is dedicated to the calculation of a class of Feynman diagrams.

\section{NRG decomposition}
\label{sec:NRG}

In this paper, we focus on stationary gravitational fields. Therefore, it is natural to perform a temporal Kaluza-Klein dimensional reduction \cite{Kol:2007bc,Kol:2011vg}
\begin{equation}
ds^{2}=g_{\mu\nu}dx^{\mu}dx^{\nu}=e^{2\phi}\left(dt-2 A_{I}dx^{I}\right)^{2}-e^{-\frac{2\phi}{d-3}}\gamma_{IJ}dx^{I}dx^{J} ~.
\label{KK}
\end{equation}
This relation defines a change of variables from $g_{\mu\nu}$ to $(\phi, A_{I}, \gamma_{IJ})$, where capital Latin indices run over the $d-1$ dimensional space, with $I, J = 1, 2, \ldots, d-1$. The new variables are referred to as ``Non-Relativistic Gravity" (NRG) fields. In fact, the parameterization of the metric in terms of NRG fields is general and can also be applied to time-dependent setups \cite{Kol:2010si}. However, it is particularly useful in the case of stationary or nearly stationary gravitational systems, see {\it e.g.,} \cite{Levi:2008nh,Levi:2011eq,Foffa:2011np,Birnholtz:2013nta,Foffa:2013gja,Levi:2015ixa,Foffa:2016rgu,Levi:2018nxp}. \footnote{The parametrization \reef{KK} is similar to, but distinct from, the ADM decomposition \cite{Kol:2010ze}.} 

For time-independent fields, the Einstein-Hilbert action takes the form 
\bea
S_\mt{EH}&=&-{1\over 16\pi} \int dt \, d^{d-1}x \sqrt{|g|} \, R[g] = 
\int dt \, S_\mt{NRG}\left(\phi,A,\gamma\right) ~,
\non
S_\mt{NRG}\left(\phi,A,\gamma\right)&=&-\frac{1}{16\pi}\int d^{d-1}x\,\sqrt{\gamma}\Big(- \mathcal{R}\left[\gamma\right]+\frac{d-2}{d-3}\partial_{I}\phi \, \partial^{I}\phi- e^{2\frac{d-2}{d-3}\phi}F_{IJ}F^{IJ}\Big)~,
\non
\label{EHaction}
\eea
where the spatial metric $\gamma_{IJ}$ is used to raise and lower the indices, $F_{IJ}=\partial_{I}A_{J}-\partial_{J}A_{I}$ and $\mathcal{R}\left[\gamma\right]$ is the Ricci scalar of the metric $\gamma_{IJ}$. Note that the integral in the definition of $S_\mt{NRG}$ is $(d-1)$- dimensional because time factors out and can be stripped off. In particular, time-independent geometries satisfy the following equations of motion
\bea
 &&\mathcal{R}_{IJ} - {\gamma_{IJ} \over 2} \mathcal{R} =  {d-2\over d-3} T^\phi_{IJ} -2 \, e^{2\frac{d-2}{d-3}\phi} T^\mt{EM}_{IJ}~,
 \non
 &&   {1\over \sqrt{\gamma}}\del_I \( \sqrt{\gamma} \, \gamma^{IJ} \,\del_J\phi\) + e^{2\frac{d-2}{d-3}\phi} F^2 =0~,
 \label{EOM}
 \\
 && \nabla_I \(e^{2\frac{d-2}{d-3}\phi} F^{IJ}\) =0 ~,
 \nonumber
\eea
where $\nabla_I$ denotes the covariant derivative compatible with $\gamma_{IJ}$, and the stress-energy tensors are 
\bea
    T_{IJ}^{\phi}&=&\partial_{I}\phi\partial_{J}\phi - \frac{\gamma_{IJ}}{2}\left(\partial\phi\right)^{2} ~,
\non
    T_{IJ}^\mt{EM}&=&F_{IA}F_{JB}\gamma^{AB} 
    - \gamma_{IJ} \Big( {1\over 4} F^2 \Big) ~.
\eea
For static geometries, $A_I$ vanishes. An example of such a geometry is provided by the Schwarzschild black hole metric in a general number of space-time dimensions, commonly known as the Tangherlini solution \cite{Tangherlini:1963bw},
\begin{equation}
ds^{2}=f\left(r\right)dt^{2}-\frac{1}{f\left(r\right)}dr^{2}-r^{2}d\Omega_{d-2}^{2} \quad , \quad 
f\left(r\right)=1-\left(\frac{r_{s}}{r}\right)^{d-3} ~,
\label{Schw}
\end{equation}
where $r_{s}$ is the Schwarzschild radius. We denote the NRG fields that describe the Schwarzschild black hole by $\phi_S, A_I^S$ and $\gamma^S_{IJ}$. Using \reef{KK}, one can read off their explicit expressions 
\be
 \phi_S= {1\over 2} \log f(r) ~, \quad A_I^S=0 ~, \quad  \gamma^S_{rr}=f^{-{d-4\over d-3}}~, \quad \gamma^S_{ij} = f^{1\over d-3} r^2 \, \Omega_{ij}~.
 \label{Sch metric}
\ee
Here and in the following sections, lower-case Latin indices run over the $(d-2)$-dimensional sphere, and $\Omega_{ij}$ represents the metric on a unit sphere. In the next section, we employ the NRG decomposition to explore gravito-magnetic perturbations of the Schwarzschild black hole.

\section{Gravito-magnetic perturbations of black hole}
\label{sec:MagLove}

In this section, we delve into the linear response of the Tangherlini solution \reef{Schw} to stationary gravito-magnetic perturbations. We determine the corresponding Love numbers without relying on the Regge-Wheeler-Zerilli equation \cite{Regge:1957td,Zerilli:1970se,Zerilli:1970wzz} and its generalization \cite{Kodama:2003jz,Ishibashi:2003ap}. The master field governed by the Regge-Wheeler equation is replaced with the NRG vector potential $A_I$. We argue that at leading order, $A_I$ decouples from other fields in the action, and its radial profile satisfies a hypergeometric equation.

\subsection{Full General Relativity analysis}
It is convenient to rewrite the equation of motion for $A_I$ in \reef{EOM} as follows
\be
 \del_I \(\sqrt{\gamma} \, e^{2\frac{d-2}{d-3}\phi} \gamma^{IK}\gamma^{JL}F_{KL}\)=0~.
 \label{eomA}
\ee
This equation is already linear in $A_I\ll 1$, allowing us to replace $\gamma_{IJ}$ and $\phi$ with the unperturbed values of the metric given by \reef{Sch metric}. Consequently, the equation of motion for $A_I$ decouples from other perturbations at linear order. In what follows, we use \reef{eomA} to explore the gravito-magnetic perturbations instead of the master field of the Regge-Wheeler equation.
 
Substituting the Schwarzschild expressions \reef{Sch metric} for $\gamma_{IJ}^S$ and $\phi_S$ into \reef{eomA} and subsequently decomposing the indices into radial and spherical components yields\footnote{Recall that the lower-case Latin indices run over the $(d-2)$-dimensional unit sphere.}
\bea
0 &=& \del_i\(f^2 \, r^{d-4} \sqrt{\Omega} \, \Omega^{ik} \, F_{kr}\)~,
\label{eomAr}
\\
 0 &=& \del_r\(f^2 \, r^{d-4} \sqrt{\Omega} \, \Omega^{j l} \, F_{r l}\)
 + \del_i\(f \, r^{d-6} \, \sqrt{\Omega} \, \Omega^{ik} \, \Omega^{j l} \, F_{k l} \) ~.
 \label{eomAsphere}
\eea
Imposing the following gauge condition for the perturbation  
\be
A_r=0 ~,
\ee
the radial equation \reef{eomAr} then takes the form 
\be
\nabla^{\mathbb{S}^{d-2}}_i \, A^i = F(\Omega) ~, \quad A^i=\Omega^{ik}A_k~.
\ee
Here, $F(\Omega)$ is an arbitrary function on a sphere, representing the residual gauge, subject to the condition $\int_{\mathbb{S}^{d-2}} F(\Omega)=0$, and $\nabla^{\mathbb{S}^{d-2}}$ denotes the covariant derivative on a unit sphere. We choose $F(\Omega)=0$. Hence, our gauge conditions read
\be
A_r=0 ~, \quad \nabla^{\mathbb{S}^{d-2}}_i \, A^i =0 ~.
\label{SchwGF}
\ee
In this gauge, the equations of motion along the sphere, \reef{eomAsphere}, are given by
\be
0=\del_r\(f^2 \, r^{d-4} \del_r A^j\)
 + f \, r^{d-6} \nabla^{\,\mathbb{S}^{d-2}}_k F^{kj}  ~,
\ee
Alternatively, employing the commutator of covariant derivatives in the second term, $\big[\nabla^{\,\mathbb{S}^{d-2}}_k, \nabla^{\,\mathbb{S}^{d-2}}_j\big] A^k=(d-3) A_j$, and using the gauge condition \reef{SchwGF} leads to
\be
 0=f \, \del_r^2 A^j+ \Big((d-4)+(d-2) \(\frac{r_{s}}{r}\)^{d-3}\Big){1\over r}\del_r A^j + (\Delta_{\mathbb{S}^{d-2}} +3-d) {1\over r^2} A^j~.
\ee
where $\Delta_{\mathbb{S}^{d-2}}$ represents the covariant Laplacian of a vector field on a unit sphere. 

Next, we introduce a dimensionless radial coordinate and separate variables 
\be
 X = \(\frac{r_{s}}{r}\)^{d-3}~, \quad A_i(X,\Omega) = X^{-{ l+1\over d-3}} A(X) Y_{i}^{lm}(\Omega) ~,
\ee
where $Y_{i}^{lm}(\Omega)$ is a divergence-free vector spherical harmonic on ${\mathbb{S}^{d-2}}$ (to satisfy the equation on the right of \eqref{SchwGF}), which is an eigenvector of the vector Laplacian on a unit sphere\footnote{In higher dimensions, $m$ is a multi-index $m=m_1...m_{d-3}$. For a detailed discussion of spherical harmonics on ${\mathbb{S}^{d-2}}$, see \cite{Higuchi:1986}.}
\bea
\nabla_{\mathbb{S}^{d-2}}^{i}Y_i^{lm}&=&0 ~,
\non
\Delta_{\mathbb{S}^{d-2}}Y_{i}^{lm}&=&\big(- l\left( l+d-3\right)+1\big)Y_{i}^{lm}~.
\eea
The equations of motion for $A^j$ reduce to a single ODE for a radial profile,
\be
 X(1-X) A''(X) + 2\((\hat l-1)X - \hat l\) A'(X)
 +\Big({d-2\over (d-3)^2} - \hat l(\hat l-1) \Big) A(X)=0 ~,
\ee
where $\hat l=l/(d-3)$, and we used
\be
\quad r\del_r =-(d-3) X\del_X~, \quad r^2\del_r^2=(d-3)^2 \, X^2 \, \del_X^2+(d-3)(d-2)X\del_X ~.
\ee
The regular solution at the horizon is given by
\be
 A(X)\propto \;{}_2F_1\Big(~1 - \hat{l}_-~,~ -\hat{l}_+~,~2~;1-X\Big)~, \quad 
 \hat{l}_\pm={l\pm 1\over d-3} ~.
\label{def:l_pm}
\ee
Expanding it around $X=0$ leads to
\be
 A_i(r,\Omega) \propto r^{ l+1}\Big(1 + \,{B(2+\hat{l}_+ \,,\hat{l}_- \,) \over B(-\hat{l}_+\,, \, -\hat{l}_-)}  \({r_s\over r}\)^{2 l+d-3} +\ldots \Big)Y_{i}^{lm}(\Omega)~,
  \label{sol_for_A}
\ee
where ellipses encode terms that include either integer or higher-order fractional powers of $X$, and $B(x,y)$ is the beta function,
\be
 B(x,y) ={ \Gamma( x) \, \Gamma( y) \over \Gamma(x+y) } ~.
\ee

\subsection{Black hole effective action}

Consider a black hole of size $r_s$ moving in a background with a typical length scale $L\gg r_s$. Given the hierarchy of scales, it is natural to employ the point particle approximation. However, this approximation has limitations, notably in its disregard for most of the physical properties of the object, particularly the finite-size effects. To overcome these limitations, one adopts an effective field theory approach \cite{Goldberger:2004jt}. Within this framework, the full GR action gives way to an effective action for the black hole, wherein the finite-size effects and internal degrees of freedom are encapsulated by an infinite tower of generally covariant terms supported at the worldline of a point-like black hole,
\be
 S_\text{p.p.}=-m\int d\tau +
 {C_{E}\over 4} \int d\tau \, E_{\mu\nu}E^{\mu\nu} + {C_{B}\over 4} \int d\tau \, B_{\alpha_1\cdots\alpha_{d-2}}B^{\alpha_1\cdots\alpha_{d-2}} + \ldots ~,
 \label{Spp}
\ee
where $d\tau$ is the proper time of the point-like black hole moving along the worldline trajectory $x(\tau)$, and
\begin{flalign}
&E_{\mu\nu}=\mathcal{R}_{\mu\alpha\nu\beta}\dot{x}^{\alpha}\dot{x}^{\beta} ~, \quad \dot{x}^{\alpha} ={dx^\alpha\over d\tau} ~,
\\&B_{\alpha_1\cdots\alpha_{d-2}}=\frac{1}{\left(d-2\right)!}\dot{x}^{\sigma}\dot{x}^{\rho}\varepsilon_{\sigma\alpha_{1}...\alpha_{d-3}\mu\nu}\mathcal{R}_{\quad \rho\alpha_{d-2}}^{\mu\nu} ~,
\end{flalign}
are the gravito-electric and gravito-magnetic components of the Riemann tensor.\footnote{Some authors use the Weyl tensor instead of the Riemann tensor in the definition of $E_{\mu\nu}$
and $B_{\alpha_1\cdots\alpha_{d-2}}$. However, the two definitions are identical for Ricci-flat geometries considered in this paper. This is because the difference between the Weyl and Riemann tensors is proportional to a linear combination of Ricci tensor and scalar curvature, which corresponds to a redundant term. In other words, it can be eliminated from the effective action through the use of field redefinition \cite{Goldberger:2007hy}.}

The coefficients $C_{E}$ and $C_{B}$ correspond to the electric-type and magnetic-type quadrupole susceptibilities of the black hole, respectively. They represent the leading finite-size effects of a spinless black hole. Notably, the gravito-electric and gravito-magnetic components of the Riemann tensor are characterized  by the angular momentum with $ l=2$, whereas tensors with higher angular momentum can be constructed by applying covariant derivatives to $E_{\mu\nu}$ and $B_{\alpha_1\cdots\alpha_{d-2}}$. Multiplying and contracting these tensors in diverse ways results in higher order finite-size effects, which represent interactions involving multipoles with different values of $ l$.

The numerical values of $C_{E}$ and $C_{B}$ are determined by matching the physical observables calculated within the EFT approach with their counterparts evaluated using the full GR theory.

In Appendix \ref{Derivation of worldline operators} we show that if the gravitational fields are weak, then the black hole effective action simplifies to\footnote{The relation between $C_E, C_B$ in \reef{Spp} and $C_l^E, C_l^B$ for $l=2$ is given by $$C_E=C^E_2\, , \quad C_B={(d-2)(d-2)!\over 2} \, C_2^B ~.$$}
\begin{multline}
S_\text{p.p.}=-m\phi\left(0\right) - {m\over 2} \phi^2(0) +\sum_{l=2}^{\infty}\frac{C_{l}^{E}}{2l!}\partial_{I_{1}}...\partial_{I_{l}}\phi\left(0\right)\partial^{I_{1}}...\partial^{I_{l}}\phi\left(0\right)\\
+\sum_{l=2}^{\infty}\frac{C_{l}^{B}}{2 l!}\partial_{I_{1}}...\partial_{I_{l-1}}F_{MN}\left(0\right)\partial^{I_{1}}...\partial^{I_{l-1}}F^{MN}(0) + \ldots ~,
\label{s_p.p}
\end{multline}
where ellipsis encodes terms that are cubic in the weak gravitational fields, and the coefficients $C_{l}^{E}$ and $C_{l}^{B}$ are referred to as Love numbers for the electric-type and magnetic-type multipole susceptibilities of the black hole.

For completeness, we include here the values of the electric-type Love numbers, $C_{l}^{E}$ as computed in 
\cite{Kol:2011vg,Cardoso:2019vof,Hui:2020xxx},
\begin{equation}
C_{l}^{E}=\frac{\Omega_{d-3+2l} }{(2\pi)^l }~
\frac{(d-2) (l+d-2)}{(d-3)(l-1)} ~
\frac{16^{\,\hat l-1}\,B\big(\,\hat l \,,\, \hat l+2\big)}{B\big( \, \hat l+\frac{1}{2} \, , \, \hat l+\frac{3}{2}\big)} ~
\tan \big(\pi \hat l\, \big) \, r_{s}^{d-3+2l} ~,
\end{equation}
where $\Omega_d$ represents the area of a unit sphere in $d$-dimensional space,
\be
\Omega_{d}={2\pi^{d\over 2} \over \Gamma\big({d\over 2}\big)}~.\label{omega}
\ee

The full EFT action consists of $S_\text{p.p.}$ and an additional term, denoted as $S_\text{bulk}$, which governs the dynamics of the long-wavelength modes of the metric. Due to general covariance, $S_\text{bulk}$ takes the same form as \reef{EHaction}, except the fields far away from the black hole are weak. Hence, we can expand around a flat metric,
\bea
S_\text{bulk}&=&-\frac{1}{16\pi}\int d^{d-1}x\,\Big[~\frac{1}{4}\partial_{K}\sigma_{IJ}\partial^{K}\sigma^{IJ}-\frac{1}{2}\partial_{I}\sigma^{IK}\partial^{J}\sigma_{JK}-\frac{1}{4}\partial_{K}\sigma\partial^{K}\sigma+\frac{1}{2}\partial_{I}\sigma^{IJ}\partial_{J}\sigma
\non
&+&\frac{d-2}{d-3}\left(\partial\phi\right)^{2}
-F_{IJ}F^{IJ}
+ \ldots \Big]~,
\label{Sbulk}
\eea
where $\sigma_{IJ}=\gamma_{IJ}-\delta_{IJ}\ll 1$, $\sigma=\delta^{IJ}\sigma_{IJ}$, and ellipsis encode self-interactions of the NRG fields. 

By construction, the EFT action is gauge invariant, so it needs to be supplemented with the gauge-fixing term $S_\mt{GF}$. We choose the harmonic gauge,\footnote{This is the square of the well-known harmonic gauge,  $g^{\mu\nu}\Gamma^\alpha_{\mu\nu}=0$, expressed in terms of NRG fields \reef{KK} and linearized.}
\begin{equation}
S_\mt{GF}=\frac{1}{32\pi}\int d^{d-1}x \, \Big[ ~ 4\(\partial_{J}A^{J}\)^2-\Big(\frac{1}{2}\partial_{J}\sigma - \partial_{K}\sigma^{K}_J\Big)^2 ~\Big] ~.
\label{GFterm}
\end{equation}
The full EFT action is given by
\be
S_\mt{EFT}=S_\text{bulk} + S_\mt{GF} + S_\text{p.p.} ~.
\ee
In the next subsection, we introduce a weak gravito-magnetic perturbation to the Schwarzschild black hole and use $S_\mt{EFT}$ to perturbatively evaluate the asymptotic value of $A_I$. Matching it to \reef{sol_for_A} allows us to determine $C_l^B$ in a general number of dimensions.

\subsection{Matching gravito-magnetic Love}

Consider a background metric of the form
\begin{equation}
\overline{A}_{I}:=C_{I J_{1} \cdots J_{l}}x^{J_{1}}...x^{J_{l}} ~, \quad \overline{\phi}=0~, \quad \overline\gamma_{IJ}=\delta_{IJ} ~,
\label{backgrnd}
\end{equation}
where $C_{I J_{1} \cdots J_{l}}$ is symmetric in all the last $l$ indices, totally traceless, and obeys the condition
\begin{equation}
C_{I J_{1} \cdots J_{l}}+C_{J_{l}I J_1...J_{l-1}}+...+C_{J_{1}...J_{l}I}=0~.
\label{cyclic}
\end{equation}
This background not only solves the linearized Einstein equations \reef{EOM} but also satisfies the gauge condition \reef{SchwGF},\footnote{The linearization of the Einstein equations is justified in a sufficiently small neighborhood around the origin where $\overline{A}_{I}\ll 1$. Moreover, $$\overline{A}_{r}={\del x^I\over \del r} \, \overline{A}_I={x^I\over r} \, C_{I J_{1},..,J_{l}} x^{J_{1}}...x^{J_{l}}=0 ~,$$ where the last equality follows from \reef{cyclic}. Finally, $\nabla^{\mathbb{S}^{d-2}}_i \, \overline A^i =0$ follows from $\del_I \overline A^I=0$ written in spherical coordinates with $\overline A_r=0$.} 
\begin{equation}
\partial_{I}\overline{F}^{IJ}=0~, \quad \overline A_r=0 ~, \quad \nabla^{\mathbb{S}^{d-2}}_i \, \overline A^i =0 ~.
\end{equation}
The total angular momentum carried by $\overline{A}_{i}$ is characterized by $l$. Furthermore, by construction $C_{I J_{1} \cdots J_{l}}$ belongs to an irreducible representation of the permutation group. Assuming $C_{I J_{1} \cdots J_{l}} r_s^l \ll 1$, we can treat this background as a small perturbation of the Schwarzschild black hole located at the origin.

To systematically evaluate perturbative gravito-magnetic corrections to the black hole's metric in the region far away from the event horizon, one needs to redefine the vector potential $A_i \to \overline A_I + A_I$ in the effective action $S_\mt{EFT}$ and subsequently calculate Feynman diagrams with one external propagator of $A_I$ and an appropriate number of $\overline A_I$ insertions.

\begin{figure}
  \centering
  \includegraphics[width=0.6\linewidth]{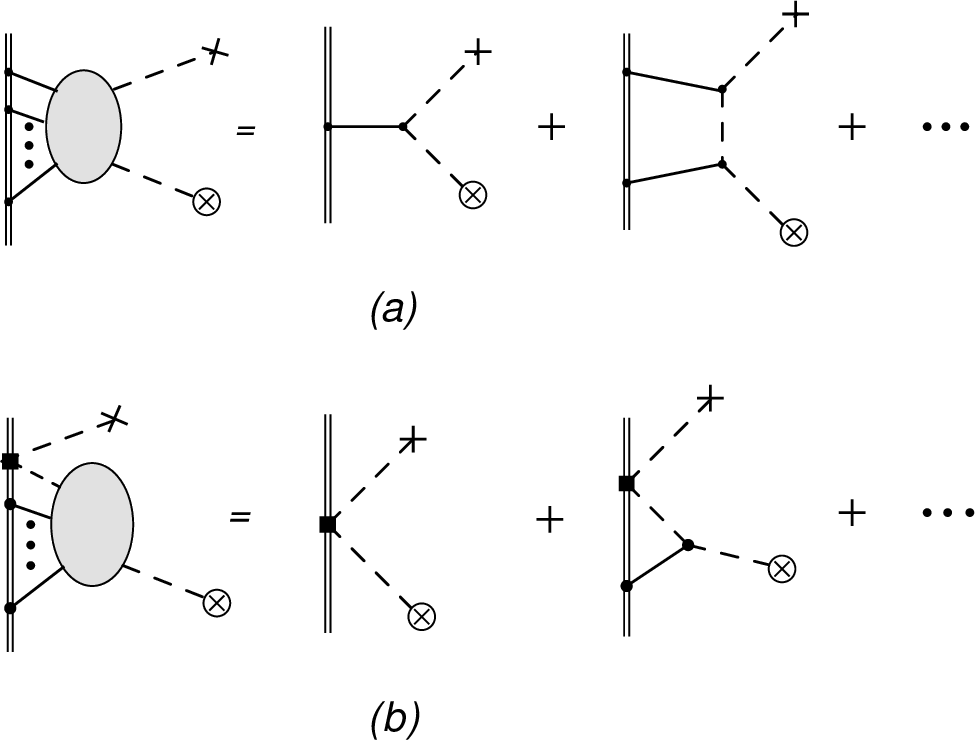}
  \caption{A class of Feynman diagrams illustrating the linear response of the Schwarzschild black hole to a weak gravito-magnetic perturbation, $\overline A_I$ (denoted by a cross in the diagrams). Solid and dashed lines represent the propagators of $\phi$ and $A_I$, respectively, while the black hole's worldline is indicated by a double line. The mass vertex is denoted by a solid dot on the worldline, and the finite-size effect associated with $C_l^B$ is represented by a solid box. Panel (a) displays a family of diagrams with only mass vertices on the worldline, and panel (b) shows a family of Feynman diagrams with a single insertion of the magnetic-type Love number $C_l^B$.}
  \label{fig:gravito-A-massDiag}
\end{figure}

In particular, the linear response of the Schwarzschild black hole induced by an external perturbation $\overline A_I$ is associated with a class of Feynman diagrams featuring a single insertion of $\overline A_I$. This class can be split into two families. The first family is shown in Fig.\ref{fig:gravito-A-massDiag}(a). The diagrams within this family carry only mass vertices on the worldline, and by dimensional analysis, they are proportional to  $\({m\over r^{d-3}}\)^n \overline A_I \sim \({r_s\over r}\)^{n(d-3)} \overline A_I$, where $n$ is an integer. The second family is depicted in Fig.\ref{fig:gravito-A-massDiag}(b), containing, in addition to the mass terms on the worldline, a single insertion of the finite-size effect associated with $C_l^B$. Consequently, the diagrams of this family are proportional to ${C_l^B\over r^{d-3+2l}} \({r_s\over r}\)^{n(d-3)} \overline A_I$.

Thus, we deduce that the diagrams belonging to the first family recover integer powers of $X=\left(\frac{r_s}{r}\right)^{d-3}$ within parentheses in \reef{sol_for_A}. In contrast, fractional powers of $X$ in \reef{sol_for_A} indicate the presence of the gravito-magnetic finite-size effect, they correspond to the diagrams in Fig.\ref{fig:gravito-A-massDiag}(b). 

\begin{figure}
  \centering
  \includegraphics[width=0.08\linewidth]{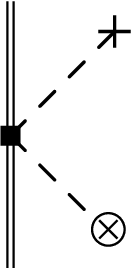}
  \caption{A Feynman diagram illustrating the black hole's response to a gravito-magnetic perturbation, $\overline{A}_I$ (represented by a cross). The black hole's worldline is denoted by a double line, while $C_l^B$ is symbolized by a solid box. The dashed line represents the propagator of $A_I$.}
  \label{fig:LovMag}
\end{figure}

To determine the value of $C_l^B$, it is sufficient to evaluate the diagram shown in Fig.\ref{fig:LovMag} and match it to the leading term of the fractional power series in \reef{sol_for_A}. We have,
\be
 \text{Fig.}\ref{fig:LovMag}=-2\,\frac{C_{l}^{B}}{ l!}\partial_{I_{1}}...\partial_{I_{l-1}}\overline F^{\,MN}(0) \left\langle  \, \partial^{I_{1}}...\partial^{I_{l-1}} \del_M A_N(0) \, A_I(x)\right\rangle ~,
\ee
where $\partial_{I_{1}}...\partial_{I_{l-1}} \overline F_{MN}(x)=2 l! \, C_{[NM] I_1 \cdots I_{l-1}}$.
The propagator of the gravito-magnetic vector potential is essentially the inverse of the quadratic in $A_I$ term in the effective action $S_\mt{EFT}$,
\be
\left\langle A_I(x)A_J(0)\right\rangle = \frac{2}{\Omega_{d-3}} \, {\delta_{IJ} \over |x|^{d-3}} ~.
\label{propA}
\ee 
Using this propagator, yields
\begin{equation}
\text{Fig.}\ref{fig:LovMag}= \, - \, \frac{4(2\pi)^l}{\Omega_{d-3+2l}} ~ \frac{(l+1)}{l } \, C_{l}^{B}\, \frac{\overline A_I}{|x|^{d-3+2l}}  ~.
\end{equation}
Hence, the EFT expression for $A_I$ is given by
\be
A_I^\mt{EFT}(x)= \overline A_I \, \Big(1 -\frac{4(2\pi)^l}{\Omega_{d-3+2l}}\frac{(l+1)}{l } \, \frac{C_{l}^{B}}{|x|^{d-3+2l}} + \ldots \Big) ~,
\label{Aeft}
\ee
where ellipses represent terms involving either integer or higher-order fractional powers of $\left(\frac{r_s}{r}\right)^{d-3}$.  
Matching this result to \reef{sol_for_A} gives
\footnote{Recall that in spherical coordinates $\overline{A}_I$ transforms to $\overline{A}_I\rightarrow\left(0,Cr^{l+1}Y_{i}^{lm}(\Omega)\right)$, where $Y^{lm}_i(\Omega)$ is the vector spherical harmonic, and $C$ is the characteristic magnitude of $C_{I A_{1}\cdots A_{l}}$.}
\be
 \boxed{C_l^B= -{\Omega_{d-3+2l}\over 8 (2\pi)^l}~
 {1+\hat{l}_+\over 1+\hat{l}_+ + \hat{l}_-} ~
 {B(\hat{l}_+,\hat{l}_-) \over B(-\hat{l}_+\,, \, -\hat{l}_-)} ~
 r_{s}^{d-3+2l}   }  ~.
 \label{magneticLove}
\ee

This determination of the gravito-magnetic Love numbers of Schwarzschild-Tangherlini black holes in all dimensions is one of the main results of this paper. Comparing with version 3 of \cite{Hui:2020xxx}, one finds that most of the expression matches, including the positions of all poles and zeros. However, there is a remaining discrepancy, a fraction which depends on $d$ and $l$, and in particular it changes the residues. Following correspondence regarding \reef{magneticLove}, \cite{Hui:2020xxx} was revised (version 4) to agree with it. 

The two real parameters, $\hat{l}_\pm$, defined in \reef{def:l_pm}, are strictly positive in the region $l\geq 2, d\geq 4$. Consequently, the beta function $B(\hat{l}_+,\hat{l}_-)$ in \reef{magneticLove} is smooth and positive in this range. In contrast, $B(-\hat{l}_+\,, \, -\hat{l}_-)$ exhibits simple poles when one or both of $\hat{l}_\pm$ equal an integer and vanishes for non-integer $\hat{l}_\pm$ if their sum is an integer. This leads to three types of $C_l^B$: finite, divergent, or vanishing. The specific type of magnetic Love number is entirely determined by the values of $\hat{l}_\pm$. If both parameters, along with their sum, are non-integers, the Love number in question is finite. If both are non-integers, while their sum is a positive integer, then the Love number diverges. Finally, if at least one of the two parameters is a positive integer, the Love number vanishes. 

For instance, in $d=6$, Love numbers satisfying $l\, \text{mod} \, 3 = 0$ diverge, whereas the remaining Love numbers vanish. In $d=5$, Love numbers with odd $l$ vanish, and those with even $l$ diverge. Notably, all magnetic Love numbers vanish in $d=4$.

\subsection{Divergent Love numbers}
\label{spacial case magnetic}

As pointed out in the previous subsection, the magnetic Love numbers \reef{magneticLove} diverge if $\hat{l}_\pm$ are non-integers, while their sum is a positive integer. In contrast, physical observables are finite. Therefore, divergent $C_l^B$ requires interpretation along with careful treatment to isolate a finite number that characterizes the physical polarization of the black hole. In this subsection, we argue that these divergences are classical analogs of the counterterms ubiquitous in the context of quantum field theory. In particular, to remedy this pathology, one follows the standard guidelines of regularization and renormalization. We illustrate that the physical magnetic Love numbers are finite and exhibit {\it classical} Renormalization Group (RG) flow when $C_l^B$ in \reef{magneticLove} is singular.

For non-integer $\hat{l}_\pm$ such that $\hat{l}_+ + \hat{l}_-=n$, where $n$ is a positive integer, the two power series in \reef{sol_for_A} merge, resulting in a regular power series augmented by a logarithmic term. The expansion \reef{sol_for_A} is replaced with
\bea
 && A_i(r,\Omega) \propto r^{ l+1}\Big(1 +  \alpha\Big(H_{\hat{l}_+ +1}+H_{\hat{l}_-}-H_{1+n} + \log X \Big) X^{\,n+1}
 +\ldots \Big)Y_{i}^{lm}(\Omega)~,
\non
&& \alpha= { (-1)^{n+1} \over n! (n+1)!} ~
 {\Gamma(2+ \hat{l}_+ ) \, \Gamma(\hat{l}_-)\over\Gamma(-\hat{l}_+) \, \Gamma(-\hat{l}_-)}  ~,
 \label{sol_for_div_A}
\eea
where $H_n$ represents the $n$-th harmonic number.  We display only the asymptotic behaviour and the leading logarithmic fall off, with the remaining terms represented by the ellipses. 

To recover \reef{sol_for_div_A} using the EFT calculation, we must sum the diagram in Fig.\ref{fig:LovMag} with those Feynman graphs in Fig.\ref{fig:gravito-A-massDiag}(a) that have $n+1$ mass vertices on the worldline.\footnote{In the special cases that we study in this subsection, the mass dimension of $C_l^B$ satisfies $$[C_l^B]=M^{d-3+2l\over d-3}=M^{n+1}~,$$ therefore, the diagrams in Fig.\ref{fig:LovMag} and Fig.\ref{fig:gravito-A-massDiag}(a) mix.} The divergence associated with the singularity in the Love number is canceled by a similar divergence in the diagrams from Fig.\ref{fig:gravito-A-massDiag}(a), resulting in a finite expression that matches \reef{sol_for_div_A}. 

To regulate the divergence, we analytically continue the calculation to $d+\epsilon$ dimensions, where $\epsilon\ll 1$ is a small dimensional regulator. The structure of diagrams in $d+\epsilon$ dimensions that feature $n+1$ mass vertices on the worldline is given by,
\be
\text{Fig.} \ref{fig:gravito-A-massDiag}(a)\propto
{1\over \epsilon} ~ {\overline A_I \, m^{n+1}\over r^{(d+\epsilon-3)(n+1)}}  + \ldots~,
\ee
where the ellipses represent the contribution of diagrams with powers of mass either higher or lower than $n+1$. Notably, the coordinate dependence of the Feynman graphs is entirely determined by dimensional analysis, while the residue of the pole in $1/\epsilon$ is the outcome of the calculation. Here, we fix it based on the requirement that the poles in $1/\epsilon$ of Fig.\ref{fig:gravito-A-massDiag}(a) and Fig.\ref{fig:LovMag} cancel, without the need for explicit calculation. Towards the end of this subsection, we evaluate this residue in a specific example, demonstrating its agreement with the anticipated pattern of the form\footnote{In principle, one could add to this expression a finite term proportional to $m_R^{n+1}$. However, it can be eliminated by properly rescaling $L$.}
\be
\text{Fig.} \ref{fig:gravito-A-massDiag}(a) = {\alpha (d-3)\over n \, \epsilon} \, {\overline A_I \, m_R^{n+1}\over r^{(d-3)(n+1)}} \Big({L\over r}\Big)^{\epsilon(n+1)} + \ldots,
\label{massDiag}
\ee 
where 
\be
 m_R(L)=  r_s^{d-3}\Big({r_s\over L}\Big)^{\epsilon} \propto m L^{-\epsilon}  ~.
\ee
In the $\epsilon\to 0$ limit, $m_R$, up to a numerical factor, represents the mass of the black hole in a $d$-dimensional spacetime.\footnote{\label{massVSrad}The relation between the mass of a black hole and its Schwarzschild radius in a general number of dimensions is given by $$m={(d-2)\Omega_{d-3}\over 8(d-3)} \, r_s^{d-3}~.$$}

In field theory terminology, $C_l^B$ in \reef{magneticLove} is referred to as a bare coupling, and its divergence serves as a counterterm essential for canceling the above $1/\epsilon$ pole to render the gravito-magnetic amplitude, $A_I$, finite. Specifically, the bare $C_l^B$ is usually decomposed as follows
\be
C_l^B  =L^\epsilon \(C_l^\mt{\,R}(L) + C_l^\mt{\,C.T.} m_R^{n+1} \) ~, \quad C_l^\mt{\,C.T.} = 
\frac{\Omega_{d-3+2l} }{8 (2\pi)^l }  ~
 {\alpha \over \hat{l}_+}~
   {d-3\over \epsilon}  ~,
\label{bareVSren}
\ee
where $L$ is an arbitrary length scale in the EFT, $C_l^\mt{\,R}(L)$ denotes a finite renormalized coupling representing the physical polarization of the black hole, and the counterterm $C_l^\mt{\,C.T.}$ stands for the leading-order divergent term in the expansion of \reef{magneticLove} around $\epsilon=0$. Substituting $C_l^B$ into \reef{Aeft}, adding \reef{massDiag} to it, and taking the $\epsilon\to 0$ limit yields a finite expression
\be
A_i^\mt{EFT} \propto r^{l+1} \Big(1 -
\frac{8 (2\pi)^l }{\Omega_{d-3+2l} } \, \frac{\hat{l}_+}{n} \, \frac{C_{l}^{R}}{r^{(d-3)(n+1)}} + {\alpha \, m_R^{n+1}\over r^{(d-3)(n+1)}} \log \Big({L\over r}\Big)^{d-3} + \ldots \Big)
Y_{i}^{lm}(\Omega) ~.
\ee

The gravito-magnetic vector potential, $A_i^\mt{EFT}$, is independent of an arbitrary scale $L$. Therefore, the physical Love number, $C_{l}^{R}$, exhibits a {\it classical} RG flow,
\be
L{d\over dL} A_i^\mt{EFT}=0 \quad \Rightarrow \quad L{d C_{l}^{R}\over dL}=
 \frac{\Omega_{d-3+2l} }{8 (2\pi)^l } ~ {n(d-3)\over \hat{l}_+} ~
   \alpha ~ r_s^{(d-3)(n+1)}~.
\ee
Solving the RG flow equation fixes $C_{l}^{R}$ up to an additive constant. This constant is defined by matching $A_i^\mt{EFT}$ to the full solution \reef{sol_for_div_A}. Identifying $L$ with $r$, yields 
\be
C_l^R(r)= {\Omega_{d-3+2l}\over 8(2\pi)^l} \, {n\over \hat{l}_+} \, \alpha 
\Big(H_{1+n} - H_{\hat{l}_+ +1} - H_{\hat{l}_-} + (d-3) \log \big({r\over r_s}\big) \Big)\,r_s^{(d-3)(n+1)} ~.
\ee
\begin{figure}
  \centering
  \includegraphics[width=0.7\linewidth]{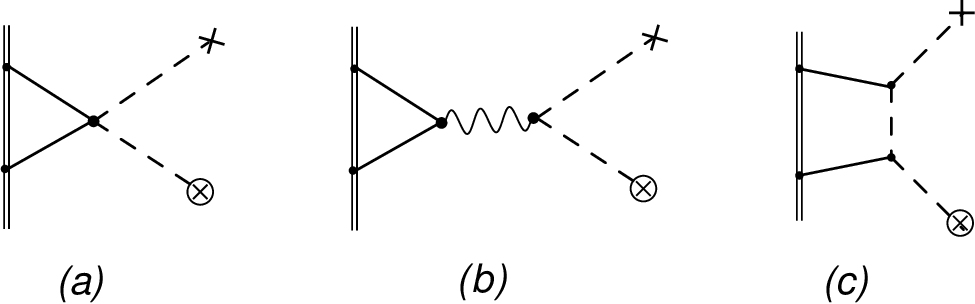}
  \caption{Feynman diagrams that mix with the graph in Fig.\ref{fig:LovMag} in the case of $d=7$ and $l=2$.  The gravito-magnetic perturbation, $\overline A_I$, is denoted by a cross. Solid, dashed and wiggly lines represent the propagators of $\phi$, $A_I$ and $\sigma_{IJ}$, respectively. The black hole's worldline is indicated by a double line, while the mass vertex is denoted by a solid dot on the worldline.}
  \label{fig:gravito-A-7D-Diag}
\end{figure}

To further illustrate the preceding discussion and provide evidence for our claims, let us examine the scenario where $d=7$ and $l=2$, corresponding to $\hat{l}_+={3\over 4}$ and $\hat{l}_-={1\over 4}$, thereby leading to $n=1$, indicating a divergence in the Love number. This case is the simplest, as it is characterized by the lowest possible values of angular momentum and the number of mass insertions on the worldline, $n+1=2$. 

The relevant Feynman diagrams featuring two mass insertions on the worldline are depicted in Fig.\ref{fig:gravito-A-7D-Diag}. We present the details of calculating these diagrams in Appendix \ref{Evaluation gravito-A-7D-Diag}. Here, we provide the final answer in general $d$:
\bea
\text{Fig.}\ref{fig:gravito-A-7D-Diag} \text{(a)}&=& \frac{3\left(d-2\right)^{2}}{2\left(d-3\right)^{2}\left(d-7\right)}
~\overline A_I \Big( {r_s\over r} \Big)^{2(d-3)}~,
\non
\text{Fig.}\ref{fig:gravito-A-7D-Diag} \text{(b)}&=&\frac{-3}{4\left(d-3\right)\left(d-5\right)\left(d-7\right)}
~\overline A_I \Big( {r_s\over r} \Big)^{2(d-3)} ~,\label{diagrams for A results}
\\
\text{Fig.}\ref{fig:gravito-A-7D-Diag} \text{(c)}&=&\frac{3(d-10)\left(d-2\right)^{2}}{8(d-3)^2(d-7)}~
\overline A_I \Big( {r_s\over r} \Big)^{2(d-3)}~.
\nonumber
\eea
As expected, these diagrams reveal a simple pole at $d=7$. The total residue matches \reef{massDiag}, and is given by
\be
\text{Fig.}\ref{fig:gravito-A-7D-Diag} = {63\over 128 (d-7)} ~
\overline A_I \, {m_R^2 \over r^8} \, \Big({L \over r}\Big)^{2\epsilon} ~.
\ee
Note that \reef{massDiag} is based on the pole structure of \reef{magneticLove}, therefore this match provides an independent confirmation of the general expression \reef{magneticLove} for the magnetic Love numbers. Furthermore, in this case we have
\be
L{d C_{2}^{R}\over dL}\Big|_{d=7} = {7\pi^2\over 1024} \, r_s^8 \quad \text{and} \quad C_{2}^R\Big|_{d=7}= {\pi^2\over 6\cdot 16^3} \Big( 84 \log 8 -185 + 168 \log \big({r\over r_s}\big) \Big)\, r_s^8~.
\ee

\section{Non-linear polarization}
\label{sec:nonlinear}

Our previous discussion focused on a subset of terms within $S_\text{p.p.}$ -- specifically those quadratic in $B_{\alpha_1\cdots\alpha_{d-2}}$ and its covariant derivatives.  These terms govern the linear response of a black hole to an external gravito-magnetic perturbation. While the linear response is dominant in the presence of a weakly curved background, it provides an incomplete description of the black hole's internal structure due to the inherently non-linear nature of General Relativity.

In addition, our analysis confirmed that $C_l^B=0$ for the Schwarzschild black hole in $d=4$ \cite{Damour:2009vw,Binnington:2009bb,Hui:2020xxx}. A similar linear behavior is exhibited by the four-dimensional Kerr black hole \cite{Pani:2015hfa,Pani:2015nua,LeTiec:2020spy,LeTiec:2020bos,Chia:2020yla,Goldberger:2020fot,Charalambous:2021mea}. Consequently, within the framework of Einstein's theory of gravity, upcoming observational data -- capturing physics beyond the point-particle approximation -- is anticipated to be shaped by non-linear effects.

To gain a comprehensive understanding of these effects beyond the linear response approximation, it is imperative to explore higher order finite-size coefficients. In this subsection, we take a step in this direction by partially unraveling the higher curvature structure of $S_\text{p.p.}$. Recent studies on this subject can be found in \cite{DeLuca:2023mio,Riva:2023rcm}.

To determine the finite-size coefficients governing non-linear effects associated with cubic and higher-order curvature terms in $S_\text{p.p.}$, one must resort to a higher-order perturbation theory. However, as we argue below, the NRG decomposition in \reef{KK} allows us to isolate a symmetry that protects certain coefficients of this type; that is, they vanish in a general number of space-time dimensions, including $d=4$.

To this end, we note that for stationary geometries, the NRG action in \reef{EHaction} exhibits manifest invariance under $A_I\to -A_I$. This symmetry is a remnant of the invariance of the full Einstein-Hilbert action under time reversal, $T$, accompanied by an appropriate transformation of the metric. In terms of NRG fields in \reef{KK}, it boils down to
\be
 T\, : \quad t \to -t ~, \quad A_I\to -A_I ~, \quad \phi \to \phi~, \quad \gamma_{IJ} \to \gamma_{IJ} ~.
 \label{T}
\ee
This orientation-reversing diffeomorphism results in a simple transformation of the {\it spatial} part of the gravito-electric and gravito-magnetic components of the Riemann tensor: $E_{IJ}$ remains unchanged under time reversal ($T$-even), while $B_{I_1\cdots I_{d-2}}$ changes sign ($T$-odd),
\be
 E_{IJ} \to E_{IJ} ~, \quad 
 B_{I_1\cdots I_{d-2}} \to -B_{I_1\cdots I_{d-2}}~.
\ee

By construction, the effective action respects all the symmetries of Einstein's theory of gravity. Consequently, one might erroneously conclude that worldline terms containing an odd number of $B_{I_1\cdots I_{d-2}}$'s are not allowed in $S_\text{p.p.}$ due to their $T$-odd nature. This is not universally true, as the coefficients multiplying curvature invariants in $S_\text{p.p.}$ are not necessarily $T$-even. These coefficients encode the internal degrees of freedom of the black hole, and their transformation may compensate for the emergent minus sign. A notable example of this kind is the coupling of a black hole's spin to a weakly curved background \cite{Porto:2006bt,Porto:2005ac,Levi:2008nh,Levi:2010zu}
\be
 \delta S_{p.p.} \propto \int d\tau \, S^{IJ} F_{IJ} ~,
\ee
where the antisymmetric tensor $S^{IJ}$ represents the spin variables of a rotating black hole. Both $S^{IJ}$ and $F_{IJ}$ change sign under time reversal, ensuring the invariance of $\delta S_{p.p.}$ under $T$.

That being said, the Schwarzschild background is static with $A^S_I=0$, rendering the metric invariant under $t\to -t$. Therefore, the finite-size coefficients of the effective action, which encode the internal structure of the Schwarzschild black hole, must be $T$-even. Thus, the upshot of our discussion is: 
\begin{tcolorbox}
The finite-size coefficients associated with worldline curvature invariants containing an odd number of $B_{I_1\cdots I_{d-2}}$'s in the effective action describing the Schwarzschild black hole are equal to zero.
\end{tcolorbox}

This conclusion becomes particularly transparent when analyzing perturbations around the Schwarzschild black hole using NRG decomposition \reef{KK}. Let us expand the NRG fields as follows
\bea
 \phi&=&\phi_S + \phi^{(1)} + \phi^{(2)} + \ldots = \phi_S + \sum_{n=1}^\infty \phi^{(n)} ~,
 \non
 \gamma_{IJ}&=& \gamma_{IJ}^S + \sigma_{IJ}^{(1)} + \sigma_{IJ}^{(2)} + \ldots = \gamma_{IJ}^S + \sum_{n=1}^\infty \sigma_{IJ}^{(n)} ~,
 \label{pert_exp}\\
 A_I&=&A_I^{(1)} + A_I^{(2)} + \ldots = \sum_{n=1}^\infty A_I^{(n)} ~,
 \nonumber
\eea
where the superscript $n$ denotes the order of the perturbation in the weak background $\overline A_I$. 

\begin{figure}
  \centering
  \includegraphics[width=0.6\linewidth]{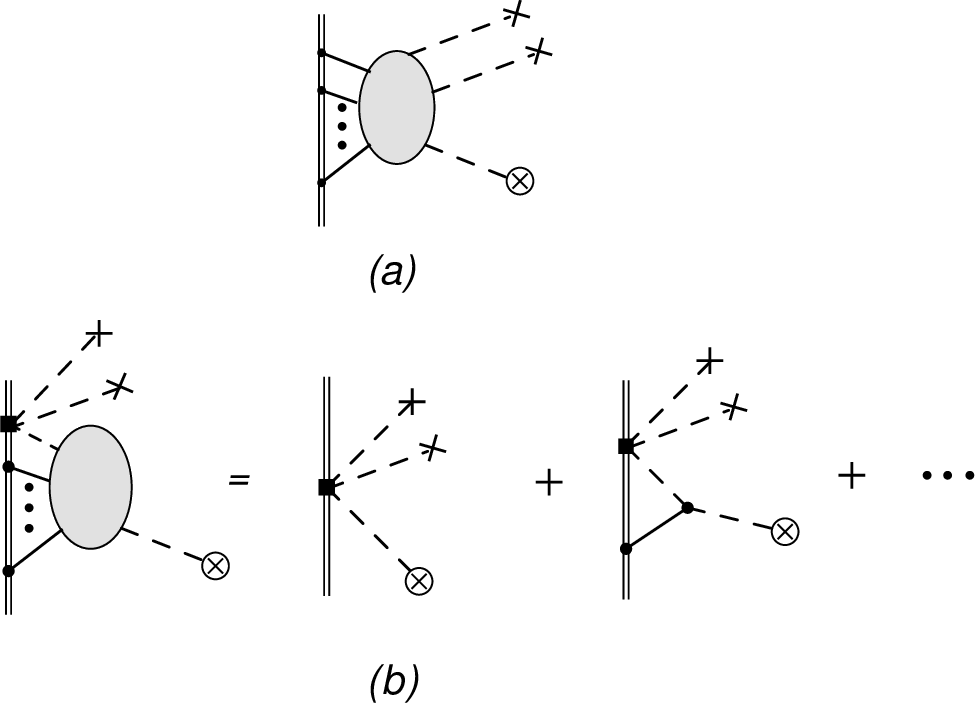}
  \caption{A class of Feynman diagrams contributing to $A_I^{(2)}$. A solid box represents the non-linear finite-size effect cubic in the gravito-magnetic component of the Riemann tensor, while the remaining Feynman rules are the same as in Fig.\ref{fig:gravito-A-massDiag}. Panel (a) displays a family of diagrams with only mass vertices on the worldline, and panel (b) shows a family of Feynman diagrams with a single insertion of the non-linear response coefficient.}
  \label{fig:NL-A-Diag}
\end{figure}

Notably, the equations of motion for $\phi$ and $\gamma_{IJ}$ in \reef{EOM} are quadratic in $A_I$. Thus, $A_I^S=0$ implies that the gravito-magnetic perturbation $\overline A_I$ leads to $\phi^{(1)} = \sigma_{IJ}^{(1)} =0$.\footnote{This no longer holds in the case of the Kerr black hole.} Consequently, the non-trivial corrections to the Schwarzschild fields $\phi_S$ and $\gamma_{IJ}^S$ begin at the second order in $\overline A_I$, resulting in $A_I^{(2)}=0$, as the equation defining $A_I$ in \reef{EOM} is already linear in the perturbation.\footnote{Alternatively, the equation for $A_I^{(2)}$ is identical to that of $A_I^{(1)}$. Therefore, $A_I^{(2)}\propto A_I^{(1)}$, indicating a trivial redefinition of the linear order amplitude.} Extending this discussion to higher orders reveals that the perturbation $\overline A_I$ gives rise to non-trivial $A_I^{(2n-1)}$ and $\phi^{(2n)}, \sigma_{IJ}^{(2n)}$ for $n \in \mathbb{N}$, whereas other corrections vanish,
\be
 A_I^{(2n)}=0 ~, \quad \phi^{(2n-1)}=0~, \quad \sigma_{IJ}^{(2n-1)}=0~, \quad \text{for} \quad  n \in \mathbb{N}~.
\ee
These constraints on the perturbative expansion \reef{pert_exp} have significant implications for the structure of $S_\text{p.p.}$: consistent with our earlier conclusion, terms with an odd number of $B_{I_1\cdots I_{d-2}}$ are protected, meaning that the corresponding finite-size coefficients vanish. 

For instance, the condition $A_I^{(2)}=0$ is equivalent to the absence of terms that are cubic in the gravito-magnetic component of the Riemann tensor. In $d=4$, this includes the coefficient of $B_{\mu \nu} B^{\nu \alpha} B_{\alpha}^{~\mu}$, which vanishes. 

In terms of NRG fields \reef{KK}, the absence of these terms in the black hole's effective action has a straightforward diagrammatic explanation. As derived from \reef{EHaction}, all bulk vertices of the effective action are either independent of $A_I$ or quadratic in $A_I$. Consequently, any $A_I$-line originating from the external leg (either $\overline A_I$ or $A_I$) continues seamlessly without branching to another external leg or terminates at the worldline vertex representing a finite-size effect. In particular, diagrams of the form shown in Fig.\ref{fig:NL-A-Diag} (a) are nonexistent, because the bulk of these diagrams has three external $A_I$-legs. The only non-trivial diagrams that contribute to $A_I^{(2)}$ necessarily include a worldline vertex representing the finite-size effect, as illustrated in Fig.\ref{fig:NL-A-Diag} (b). Matching the leading-order diagram on the right-hand side of Fig.\ref{fig:NL-A-Diag} (b) with $A_I^{(2)}=0$ leads to the vanishing of the finite-size coefficient associated with the term cubic in the gravito-magnetic component of the Riemann tensor.

Similarly, the vanishing of $\phi^{(3)}$ forbids mixing of 3 gravito-magnetic components $B_{\alpha_1\cdots\alpha_{d-2}}$ with a single $E_{\mu\nu}$.\footnote{Vanishing of $\phi^{(1)}$ excludes a term of the form $B_{\mu\nu} E^{\mu\nu}$ in $d=4$, which is a quite trivial result, because linear order perturbation theory does not mix odd and even sectors.}

\section{Discussion}
\label{sec:discussion}

In this paper, we computed the gravito-magnetic Love numbers \reef{magneticLove} for any space-time dimension. To facilitate the derivation, we decomposed the metric in terms of NRG fields \reef{KK} and calculated the Love numbers using the equation of motion for the vector field $A_I$. At linear order $A_I$ decouples from other perturbations thus providing a direct and concise way of determining the Love numbers, avoiding the use of the Ishibashi-Kodama master field. Notably, the revised version of \cite{Hui:2020xxx} agrees with the final result.

The second thrust of our paper revolves around nonlinear response coefficients. We have identified a parity-based selection rule for nonlinear terms encompassing both electric-type and magnetic-type gravitational field tensors. This implies the vanishing of numerous nonlinear Love numbers, streamlining future research efforts to concentrate on a considerably smaller set of response coefficients.

Moreover, it follows from \reef{EOM} and the analysis in Section \ref{sec:nonlinear} that considering perturbations beyond the linear response theory results in the mixing between the gravito-magnetic and gravito-electric sectors. Therefore, to capture the nonlinear effects of the polarization of the Schwarzschild black hole, it is necessary to incorporate mixed-type higher-order curvature terms in the effective action \reef{Spp}, involving both the gravito-electric and gravito-magnetic components of the Riemann tensor. For instance, the full list of mixed-type terms relevant for the second-order effects associated with perturbations characterized by $l=2$ in $d=4$ consits of two terms
\be
 C_\mt{BBE}\int d\tau B^{\mu \nu}B_{\nu}^{\alpha}E_{\alpha\mu}  \quad \text{and} \quad 
 C_\mt{EEB}\int d\tau E^{\mu \nu}E_{\nu}^{\alpha}B_{\alpha\mu}~.
\ee
As explained in Section \ref{sec:nonlinear}, the coefficient $C_\mt{EEB}$ vanishes since this term does not respect parity in \reef{T}. Consequently, only the coefficient $C_\mt{BBE}$ needs to be determined to capture second-order effects induced by perturbations with $l=2$.\footnote{The coefficient of $B^{\mu \nu}B_{\nu}^{\alpha}B_{\alpha\mu}$ vanishes because, similar to $E^{\mu \nu}E_{\nu}^{\alpha}B_{\alpha\mu}$, this term does not respect parity. The vanishing of the finite-size coefficient of $E^{\mu \nu}E_{\nu}^{\alpha}E_{\alpha\mu}$ was recently reported in \cite{Riva:2023rcm}.} In our future work, we aim to calculate this coefficient \cite{inprogress}.

 {\bf Acknowledgements}  We thank Lam Hui, Luca Santoni, Adam Solomon and especially Austin Joyce for helpful discussions and correspondence. TH and MS acknowledge partial support from an Israel Science Foundation (ISF) Center for Excellence grant (Grant Number 2289/18). MS is also partially supported by BSF Grant No. 2022113 and NSF-BSF Grant No. 2022726. BK acknowledges support by the Israel Science Foundation (grant No. 1345/21) and by a grant from Israel's Council of Higher Education. 

\appendix

\section{Finite-size terms}
\label{Derivation of worldline operators}

In this Appendix, we derive Eq. \reef{s_p.p}. This part of the complete world line action exhibits quadratic dependence on weak gravitational fields, capturing the black hole's linear response to a weakly curved background. 

The first two terms in \reef{s_p.p} follow from the weak field approximation and NRG decomposition \reef{KK} of the mass term,
\be
 -m\int d\tau = -m \int \sqrt{g_{00}} \, dt = -m \Big(1+\phi(0)+{\phi^2(0)\over 2} + \ldots\Big) \int dt ~.
\ee
In \reef{s_p.p}, we suppress the irrelevant time interval.

The full effective action maintains general covariance and reparametrization invariance. As a result, the remaining terms in \reef{s_p.p} should arise from worldline terms that are quadratic in the Riemann tensor and its covariant derivatives. Using the gravito-electric and gravito-magnetic components of the Riemann tensor, the full list of relevant terms is given by
\be
 \nabla_{\beta_{1}}...\nabla_{\beta_{l-2}}E_{\mu\nu}~, \quad \nabla_{\beta_{1}}...\nabla_{\beta_{l-2}}B_{\alpha_{1},...,\alpha_{d-2}}~, \quad  l\geq 2~,
\ee
where $ l$ characterizes the spin of the tensor. To linearize these terms, we use  
\begin{equation}
\mathcal{R}_{\mu\alpha\nu\beta}=\frac{1}{2}\left(h_{\alpha\nu,\mu\beta}+h_{\mu\beta,\alpha\nu}-h_{\alpha\beta,\mu\nu}-h_{\mu\nu,\alpha\beta}\right) + \mathcal{O}(h^2)~,
\label{Rlinear}
\end{equation}
where $g_{\mu\nu}=\eta_{\mu\nu} + h_{\mu\nu}$ is a weakly curved background with $h_{\mu\nu}\ll 1$. For a stationary case, the metric is time-independent and $\dot x^\alpha=\delta^\alpha_0$. Hence, the non-trivial components of $E_{\mu\nu}$ are given by
\begin{equation}
E_{IJ}=- \frac{1}{2} h_{00,I J} + \mathcal{O}(h^2)= -  \del_I \del_J \phi  + \mathcal{O}(h^2) ~.
\end{equation}
In particular, for a given $ l$, there exists only one term that is quadratic in the gravito-electric component of the Riemann tensor. It is represented by the last term in the first line of \reef{s_p.p}.

Similarly, the non-vanishing gravito-magnetic components of $B_{\alpha_1\cdots\alpha_{d-2}}$ satisfy
\be
B_{I_1\cdots I_{d-2}}=\frac{1}{\left(d-2\right)!}\varepsilon_{0I_{1}...I_{d-3}}^{\quad\quad\quad MN}\mathcal{R}_{MN 0I_{d-2}} ~.
\ee
Using the linearized version of the Riemann tensor, as given in \reef{Rlinear}, and neglecting temporal derivatives, results in
\bea
 B_{I_{1}...I_{d-2}}&=&\frac{1}{\left(d-2\right)!}\varepsilon_{0I_{1}...I_{d-3}}^{\quad\quad\quad MN}\frac{1}{2}
 \left(h_{N0,I_{d-2}M}-h_{M0,NI_{d-2}}\right) + \mathcal{O}(h^2)
 \nonumber \\
 &=& \frac{-1}{(d-2)!}\varepsilon_{0I_{1}...I_{d-3}}^{\quad\quad\quad MN} \, \del_{I_{d-2}}F_{MN}  + \mathcal{O}(h^2) ~.
\eea
To derive the worldline term in the second line of \reef{s_p.p}, one needs to apply $ l-2$ derivatives to the gravito-magnetic component and square it. This involves contracting the $2(d+ l-4)$ free indices in all possible inequivalent ways. Due to the antisymmetry of the Levi-Civita symbol, only two candidates exist for non-trivial (and inequivalent) way of contracting the indices:

\noindent
1. All free spatial indices of the Levi-Civita symbol in one gravito-magnetic component are contracted with the free indices of the Levi-Civita symbol in the other gravito-magnetic component. The remaining $ l-1$ free indices of derivatives in both components are contracted among themselves.

\noindent
2. One index of the Levi-Civita symbol is contracted with one of the derivatives present in the other gravito-magnetic component.

However, using the following identities,
\bea
\varepsilon_{0I_{1}...I_{d-3}\, MN}\,\varepsilon_{0I_{1}...I_{d-3} AB} &=&
(d-3)!\(\delta_{MA}\delta_{NB}-\delta_{MB}\delta_{NA}\) ~,
\\
\varepsilon_{0I_{1}...I_{d-4}\, I MN}\,\varepsilon_{0I_{1}...I_{d-4}JAB} &=& (d-4)!\[ \delta_{IJ}\(\delta_{MA}\delta_{NB}-\delta_{MB}\delta_{NA}\) - \(J \leftrightarrow A\, , ~J \leftrightarrow B\) \] ~,
\nonumber
\eea
it is straightforward to demonstrate that both alternatives of contracting the indices lead to the same term displayed in the second line of \reef{s_p.p}.\footnote{Note that terms proportional to the equations of motion, $\del_I F^{IJ}$, are redundant and should be ignored in the effective action.}

\section{Evaluation of diagrams in Fig.\ref{fig:gravito-A-7D-Diag} }
\label{Evaluation gravito-A-7D-Diag}

In this appendix, we provide a detailed explanation of how to evaluate the diagrams in Fig. \ref{fig:gravito-A-7D-Diag}. We begin by supplementing the effective action \reef{Sbulk} with interaction terms that are cubic and quartic in the weak NRG fields. Expanding the NRG action \reef{EHaction} results in
\begin{multline}
  S_{\text{bulk}}\supseteq -\frac{1}{16\pi}\frac{d-2}{d-3}\int d^{d-1}x \(
  \left(\frac{\sigma}{2}\delta^{IJ}-\sigma^{IJ}\right)\partial_{I}\phi\partial_{J}\phi-2\phi F_{IJ}F^{IJ}-2\frac{d-2}{d-3}\phi^{2}F_{IJ}F^{IJ}\)\\
  -\frac{1}{16\pi}\int d^{d-1}x ~ \sigma_{K}^{J}\left(2F_{IJ}F^{IK}-\frac{1}{2}F_{MN}F^{MN}\delta_{J}^{K}\right)+\ldots~,
\end{multline}
where the ellipsis encode interaction terms that are irrelevant for our purpose. The above terms give rise to the following Feynman rules in position space:\footnote{Our sign convention for the interaction vertices is consistent with the choice of sign for the propagators. The latter is completely fixed by inverting the quadratic part of the action \eqref{Sbulk}.}

\vspace{1cm}
\begin{tikzpicture}
    \begin{feynman} [scale=1.0, transform shape]
        
        \vertex  (b);
        \vertex [below=of b] (c);
        \vertex [right=1.5cm of $(b)!0.5!(c)$] (e);
        \vertex [right=of e] (f);
        
        \diagram* {
            (b) -- [very thick] (e),
            (c) -- [very thick] (e),
            (e) -- [photon] (f),
        };
    
        \vertex [dot, minimum size=3pt] at (e) {};
        \node [right, yshift=0mm] at ([xshift=0.7cm]f) {=\quad$\frac{1}{16\pi}\frac{d-2}{d-3}\int d^{d-1}x\left(\frac{\sigma}{2}\delta^{IJ}-\sigma^{IJ}\right)\partial_{I}\phi\partial_{J}\phi ~,$};
    \end{feynman}
\end{tikzpicture}

\vspace{1cm}
\begin{tikzpicture}
    \begin{feynman} [scale=1.0, transform shape]
        
        \vertex  (b);
        \vertex [below=of b] (c);
        \vertex [right=1.5cm of $(b)!0.5!(c)$] (e);
        \vertex [right=of e] (f);
        
        \diagram* {
            (b) -- [dashed] (e),
            (c) -- [dashed] (e),
            (e) -- [very thick] (f),
        };
    
        \vertex [dot, minimum size=3pt] at (e) {};
        \node [right, yshift=0mm] at ([xshift=0.7cm]f) {=\quad$-\frac{1}{8\pi}\frac{d-2}{d-3}\int d^{d-1}x\phi F_{IJ}F^{IJ}~,$};
    \end{feynman}
\end{tikzpicture}

\vspace{1cm}
\begin{tikzpicture}
    \begin{feynman} [scale=1.0, transform shape]
        
        \vertex  (b);
        \vertex [below=of b] (c);
        \vertex [right=1.5cm of $(b)!0.5!(c)$] (e);
        \vertex [right=of e] (f);
        
        \diagram* {
            (b) -- [dashed] (e),
            (c) -- [dashed] (e),
            (e) -- [photon] (f),
        };
    
        \vertex [dot, minimum size=3pt] at (e) {};
        \node [right, yshift=0mm] at ([xshift=0.7cm]f) {=\quad$\frac{1}{8\pi}\int d^{d-1}x\sigma_{K}^{J}\left(F_{IJ}F^{IK}-\frac{1}{4}F_{MN}F^{MN}\delta_{J}^{K}\right)~,$};
    \end{feynman}
\end{tikzpicture}

\vspace{1cm}
\begin{tikzpicture}
    \begin{feynman} [scale=1.0, transform shape]
        \vertex (a);
        \vertex [below=0.7cm of a] (b);
        \vertex [below=of b] (c);
        \vertex [below=0.7cm of c] (d);
        \vertex [right=1.5cm of $(b)!0.5!(c)$] (e);
        \vertex [right=3cm of b] (e1);
        \vertex [right=3cm of c] (e2);
        
        \diagram* {
           
            (b) -- [very thick] (e),
            (c) -- [very thick] (e),
            (e) -- [dashed] (e1),
            (e) -- [dashed] (e2),
        };
       
        \vertex [dot, minimum size=3pt] at (e) {};
        \node at ([xshift=5.1cm]e) {=\quad$-\frac{1}{8\pi}\frac{\left(d-2\right)^{2}}{\left(d-3\right)^{2}}\int d^{d-1}x\phi^{2}F_{IJ}F^{IJ}~.$};
        
    \end{feynman}
\end{tikzpicture}

\vspace{1cm}
\noindent 
Here, as in the main text, the dashed line represents the vector field $A_I$, while the solid and wiggly lines denote the $\phi$ and $\sigma_{IJ}$ fields, respectively. Their propagators are given by  \reef{propA} and\footnote{To compute the $\sigma_{IJ}$ propagator, one needs to find the inverse of the quadratic term involving $\sigma_{IJ}$ in the bulk action \reef{Sbulk}, which is supplemented with the gauge fixing term \eqref{GFterm}. This involves inverting the tensor structure of the form $\delta^{IK}\delta^{JL}+\delta^{IL}\delta^{JK}-\delta^{IJ}\delta^{KL}$ in the vector space of symmetric rank 2 tensors, resulting in $\frac{1}{4}\big(\delta_{IK}\delta_{JL}+\delta_{IL}\delta_{JK}-\frac{2}{\left(d-3\right)}\delta_{IJ}\delta_{KL}\big)$.}

\begin{align}
&\left\langle \phi\left(x\right)\phi\left(0\right)\right\rangle =-\frac{4\left(d-3\right)}{\left(d-2\right)\Omega_{d-3}|x|^{d-3}}~,
\label{propPhi}\\
&\left\langle \sigma_{IJ}\left(x\right)\sigma_{KL}(0)\right\rangle =-\frac{8}{\Omega_{d-3}|x|^{d-3}}\left(\delta_{IK}\delta_{JL}+\delta_{IL}\delta_{JK}-\frac{2}{d-3}\delta_{IJ}\delta_{KL}\right)~,
\end{align}
where $\Omega_d$ is defined in \reef{omega}. Vertices featuring the insertion of the background \reef{backgrnd} are obtained by replacing one of the field strength tensors, $F_{IJ}$, with $2\overline{F}_{IJ}$ in the above Feynman rules. Additionally, we need the worldline vertex associated with the mass of the black hole effective action in \eqref{s_p.p},

\vspace{1cm}

\begin{tikzpicture}
    \begin{feynman} [scale=1.0, transform shape]
        \vertex (a);
        \vertex [below=1cm of a] (b);
        \vertex [below=1cm of b] (c);
        \vertex [right=of b] (d);
        
        \diagram* {
            (a) -- [double,  style={/tikz/double distance=2pt}] (b),
            (b) -- [double,  style={/tikz/double distance=2pt}] (c),
            (b) -- [ very thick] (d),
        };
        \node [right, yshift=0mm] at ([xshift=0.7cm]d) {=\quad$m\phi\left(0\right)$~.};
        \vertex [dot, minimum size=2pt] at (b) {};
    \end{feynman}
\end{tikzpicture}

\vspace{1cm}\noindent
Using these rules the diagram in Fig.\ref{fig:gravito-A-7D-Diag}(a) takes the form 
\begin{multline}
\text{Fig.}\,\ref{fig:gravito-A-7D-Diag}\text{(a)} = - \frac{m^{2}}{2\pi}\frac{\left(d-2\right)^{2}}{\left(d-3\right)^{2}}\int d^{d-1}x_2\left( \left\langle \phi\left(0\right)\phi(x_2)\right\rangle^2 \, \partial^{I}\overline{A}^{J}(x_2) \, \left\langle F_{IJ}(x_2) A_{N}(x_1)\right\rangle \right)~.
\end{multline}
Substituting the background field \reef{backgrnd} and the propagator \reef{propPhi} yields
\begin{equation}
\text{Fig.}\,\ref{fig:gravito-A-7D-Diag}\text{(a)} =\frac{-1}{4\pi}\frac{\left(d-2\right)^{2}}{\left(d-3\right)^{2}}\int d^{d-1}x_2\left[\left(\frac{r_{s}}{x_2}\right)^{2\left(d-3\right)}x_2^{M}C_{[JI]M}\left\langle \partial_{I}A_{J}(x_2)A_{N}(x_1)\right\rangle \right]~,
\end{equation}
where we expressed the mass, $m$, of a black hole in terms of the Schwarzschild radius, see footnote \ref{massVSrad}. In the above expression, the derivative $\partial_{I}$ acts on $x_2$. However, since the propagator of $A_I$ depends on the difference of coordinates, we can replace $\partial^{I}_{x_2}$ with $-\partial^{I}_{x_1}$ and pull it out of the integral. Substituting \reef{propA}, we obtain
\begin{equation}
\text{Fig.}\,\ref{fig:gravito-A-7D-Diag}\text{(a)} =
\frac{-r_s^{2d-6}}{2\pi \, \Omega_{d-3}}\frac{\left(d-2\right)^{2}}{\left(d-3\right)^{2}} \, C_{[IN]M} \, \partial_{x_1}^{I} 
\int d^{d-1}x_2\frac{x_2^M}{\left(x_2^2\right)^{d-3}\left(x_{21}^2\right)^{\frac{d-3}{2}}}~, \quad x_{21}=x_2-x_1~.
\end{equation}
To carry out the integral over $x_2$, we employ the following identity, derived using the Feynman parameterization trick 
\begin{equation}
\int d^d x_3\frac{x_{32}^I}{\left(x_{32}^2\right)^{\alpha}\left(x_{31}^2\right)^{\beta}}=\pi^{\frac{d}{2}}\frac{\Gamma\left(\alpha+\beta-\frac{d}{2}\right)}{\Gamma\left(\alpha\right)\Gamma\left(\beta\right)}\frac{\Gamma\left(\frac{d}{2}+1-\alpha\right)\Gamma\left(\frac{d}{2}-\beta\right)}{\Gamma\left(d+1-\alpha-\beta\right)}\left(x_{12}^2\right)^{\frac{d}{2}-\alpha - \beta}x_{12}^I \label{integral 1}~.
\end{equation}
As a result, we get
\begin{equation}
\text{Fig.}\,\ref{fig:gravito-A-7D-Diag}\text{(a)} = r_s^{2d-6}
\frac{(d-2)^{2}}{2(d-4)(d-7)\left(d-3\right)^{2}} \, C_{[IN]M} \, \partial^{I} \({x_1^M\over (x_1^2)^{d-4}}\)~.
\end{equation}
Finally, using \eqref{cyclic} and tracelessness of $C_{IJK}$ yields the expression appearing in \eqref{diagrams for A results}.

The other two diagrams in Fig.\ref{fig:gravito-A-7D-Diag} can be evaluated by following similar steps. Instead of presenting explicit computations, we list three additional master integrals essential for evaluating the two remaining diagrams. Their derivation is straightforward, relying entirely on the Feynman parameterization trick. In all the expressions below, we continue using a shorthand notation where $x_{12}=x_1-x_2$.
\begin{equation}
\int d^d x_3\,\frac{1}{\left(x_{32}^2\right)^{\alpha}\left(x_{31}^2\right)^{\beta}}=\pi^{\frac{d}{2}}\frac{\Gamma\left(\alpha+\beta-\frac{d}{2}\right)\Gamma\left(\frac{d}{2}-\alpha\right)\Gamma\left(\frac{d}{2}-\beta\right)}{\Gamma\left(\alpha\right)\Gamma\left(\beta\right)\Gamma\left(d-\alpha-\beta\right)}\left(x_{12}^{2}\right)^{\frac{d}{2}-\alpha-\beta}\label{integral 0}~,
\end{equation}

\begin{align}
&\int d^d x\,\frac{x_{32}^I x_{32}^J}{\left(x_{32}^2\right)^{\alpha}\left(x_{31}^{2}\right)^\beta} \nonumber =\pi^{\frac{d}{2}}\frac{\Gamma\left(\alpha+\beta-1-\frac{d}{2}\right)\Gamma\left(\frac{d}{2}+1-\alpha\right)\Gamma\left(\frac{d}{2}-\beta\right)}{\Gamma\left(\alpha\right)\Gamma\left(\beta\right)\Gamma\left(d+2-\alpha-\beta\right)}\,\times 
\nonumber \\
&\qquad\left[\Big(\frac{d}{2}+1-\alpha\Big)\Big(\alpha+\beta-1-\frac{d}{2}\Big)x_{12}^I x_{12}^J+\Big(\frac{d}{2}-\beta\Big)\frac{x_{12}^2}{2}\,\delta^{IJ}\right]\left(x_{12}^2\right)^{\frac{d}{2}-\alpha-\beta}\label{integral 2} ~,
\end{align}

\begin{align}
&\int d^d x_3 \,\frac{x_{32}^I x_{32}^J x_{32}^K}{\left(x_{32}^2\right)^{\alpha} \left(x_{31}^2\right)^{\beta}} =\pi^{\frac{d}{2}}\frac{\Gamma\left(\alpha+\beta-1-\frac{d}{2}\right)\Gamma\left(\frac{d}{2}+2-\alpha\right)\Gamma\left(\frac{d}{2}-\beta\right)}{\Gamma\left(\alpha\right)\Gamma\left(\beta\right)\Gamma\left(d+3-\alpha-\beta\right)} ~ \left(x_{12}^2\right)^{\frac{d}{2}-\alpha-\beta} ~ \times 
\nonumber \\
&\left[\left(\frac{d}{2}+2-\alpha\right)\left(\alpha+\beta-1-\frac{d}{2}\right)x_{12}^I x_{12}^J x_{12}^K +\left(\frac{d}{2}-\beta\right)\frac{x_{12}^2}{2}\left(\delta^{IJ} x_{12}^K +\delta^{IK} x_{12}^J+\delta^{JK} x_{12}^I \right)\right]~.
\label{integral 3}
\end{align}

\bibliographystyle{utphys.bst}
\bibliography{Refs.bib}

\end{document}